\documentclass[11pt,twoside]{article}
\usepackage{graphicx}
\usepackage{amsmath}
\usepackage{amssymb}
\usepackage{cite}
\usepackage{mathrsfs}

\begin{document}

\newcommand{\pst}{\hspace*{1.5em}}

\newcommand{\rigmark}{\em Journal of Russian Laser Research}
\newcommand{\lemark}{\em Volume 31, Number 2, 2010}

\newcommand{\be}{\begin{equation}}
\newcommand{\ee}{\end{equation}}
\newcommand{\bm}{\boldmath}
\newcommand{\ds}{\displaystyle}
\newcommand{\bea}{\begin{eqnarray}}
\newcommand{\eea}{\end{eqnarray}}
\newcommand{\ba}{\begin{array}}
\newcommand{\ea}{\end{array}}
\newcommand{\arcsinh}{\mathop{\rm arcsinh}\nolimits}
\newcommand{\arctanh}{\mathop{\rm arctanh}\nolimits}
\newcommand{\bc}{\begin{center}}
\newcommand{\ec}{\end{center}}

\thispagestyle{plain}

\label{sh}


\begin{center} {\Large \bf
\begin{tabular}{c}
UNITARY AND NON-UNITARY MATRICES \\ AS A SOURCE OF DIFFERENT BASES \\ OF OPERATORS ACTING ON HILBERT SPACES \\[-1mm]
\end{tabular}
 } \end{center}

\bigskip

\begin{center} {\bf
Sergey N. Filippov$^{1,2}$ and Vladimir I. Man'ko$^{1,2}$ }
\end{center}

\medskip

\begin{center}
{\it
$^{1}$Moscow Institute of Physics and Technology (State University)\\
Institutskii per. 9, Dolgoprudnyi, Moscow Region 141700, Russia\\
\smallskip
$^{2}$P.~N.~Lebedev Physical Institute, Russian Academy of Sciences\\
Leninskii Prospect 53, Moscow 119991, Russia}

\smallskip

e-mail:~~~sergey.filippov~@~phystech.edu,~~~manko~@~sci.lebedev.ru
\end{center}

\begin{abstract}\noindent
Columns of $d^2 \times N$ matrices are shown to create different
sets of $N$ operators acting on $d$-dimensional Hilbert space.
This construction corresponds to a formalism of the star-product
of operator symbols. The known bases are shown to be partial cases
of generic formulas derived by using $d^2 \times N$ matrices as a
source for constructing arbitrary bases. The known examples of the
SIC-POVM, MUBs, and the phase-space description of qubit states
are considered from the viewpoint of the developed unified
approach. Star-product schemes are classified with respect to
associated $d^2\times N$ matrices. In particular, unitary matrices
correspond to self-dual schemes. Such self-dual star-product
schemes are shown to be determined by dequantizers which do not
form POVM.
\end{abstract}


\section{\label{introduction} Introduction}
\pst Spin states are usually described by spinors (pure states) or
density matrices associated with a finite-dimensional Hilbert
space. On the other hand, in the tomographic-probability
representation, spin states (qudit states) can be described by
fair probability distributions or points on the simplex
(probability
vectors)~\cite{dodonovPLA,oman'ko-jetp,serg-inverse-spin}. The
maps of qudit states onto different quasidistribution functions
defined on a finite number of points are discussed
in~\cite{vourdas04,vourdas06,wootters,klimov06,klimov07}. All
these maps including the tomographic-probability
map~\cite{serg-spin,serg-chebyshev,ibort} can be formulated in
terms of star-product schemes~\cite{oman'ko-JPA,oman'ko-vitale}.
These schemes are analogues to the known scheme developed for the
star product on a phase space~\cite{stratonovich,garcia-bondia}.

The analogues of Wigner function on a finite set of points are
studied in~\cite{chaturvedi}. Among the possible probability
descriptions of qudit states one can point out a symmetric
informationally complete (SIC) positive operator-valued measures
(POVMs) studied in~\cite{caves,renes,fuchs-2010}. These maps are
associated with the existence of specific bases in
finite-dimensional Hilbert spaces which can also be considered
from the star-product point of view~\cite{serg-sic}. Another kind
of specific bases in finite-dimensional Hilbert spaces is so
called mutually unbiased bases
(MUBs)~\cite{ivanovic,wootters87,wootters-fields,albouy}. Also,
MUBs can be considered by using the star-product approach (see,
e.g., remarks in~\cite{serg-mub}). Some experimental aspects
related to SIC-POVMs and MUBs are considered in~\cite{steinberg}.

The aim of our article is to demonstrate the possibility to
construct specific bases in finite-dimensional Hilbert spaces by
using properties of unitary and non-unitary matrices. The
$d^2\times N$ matrices are built by considering $N$ operators
acting on a $d$-dimensional Hilbert space as $d^2$-dimensional
vectors. Since each $d^2\times N$ matrix corresponds to a
star-product scheme, a classification of star-product schemes with
respect to associated $d^2\times N$ matrices is given. In
particular, unitary matrices are shown to be responsible for
self-dual schemes. It turns out that there exists no minimal
self-dual star-product scheme with dequantizers in the form of
POVM effects. Also, we prove that Hermitian dequantizers and
quantizers of a self-dual scheme must contain negative
eigenvalues.

The article is organized as follows.

In Sec. 2, we present a review of Hilbert spaces as well as
representation of matrices by vectors and vice versa. In Sec. 3,
we review a star-product scheme
following~\cite{oman'ko-JPA,oman'ko-vitale}. In Sec. 4, we relate
properties of unitary and non-unitary matrices with self-dual and
other star-product schemes. In this section, we also present
star-product picture of qubit state bases, and review the known
results of constructing the different bases for qubit states
studied in~\cite{caves,renes,fuchs-2010,livine}. The conclusions
and prospects are given in Sec. 5.

\section{\label{sec-Concise-Review-Hilbert-Spaces} Concise Review of Hilbert Spaces}
\pst We review in this Section the construction of star products
of functions of discrete variables following
\cite{oman'ko-JPA,oman'ko-vitale,ibort}.

Let $\mathcal{H}_d$ be a $d$-dimensional Hilbert space of complex
vectors $|\psi\rangle$ with a standard inner product $\langle \phi
| \psi \rangle$ that is antilinear in the first argument and
linear in the second one. The normalized vectors ($\langle \psi |
\psi\rangle = 1$) describe pure states of a $d$-dimensional
quantum system (qudit). By $\mathcal{B}(\mathcal{H}_d)$ denote a
set of linear operators acting on $\mathcal{H}_d$. Since
$\dim\mathcal{H}_d=d<\infty$, any operator
$\hat{A}\in\mathcal{B}(\mathcal{H}_d)$ is bounded and thoroughly
described by the $d\times d$ matrix $A$ with complex matrix
elements $A_{ij} = \langle e_i | \hat{A} | e_i \rangle = {\rm
Tr}\big[ \hat{E}_{(i,j)}^{\dag} \hat{A} \big]$, where
$\{|e_k\rangle\}_{k=1}^{d^2}$ is an orthonormal basis in
$\mathcal{H}_d$ and $\hat{E}_{(i,j)} = | e_i \rangle \langle e_j
|$ is a matrix unit. We have just introduced the inner product of
operators $\hat{X}$ and $\hat{Y}$ in the following manner ${\rm
Tr}\big[ \hat{X}^{\dag} \hat{Y} \big] \equiv {\rm Tr}\big[
{X}^{\dag} {Y} \big]$, where matrix $X^{\dag} = (X^{\ast})^{\rm
tr} = (X^{\rm tr})^{\ast}$ determines the adjoint operator
$\hat{X}^{\dag}$. Matrix units $\hat{E}_{(i,j)}$, $1 \le i,j \le
d$, form a bases in $\mathcal{B}(\mathcal{H}_d)$. The above
arguments allow drawing a conclusion that
$\mathcal{B}(\mathcal{H}_d)$ is the $d^2$-dimensional Hilbert
space.

\subsection{\label{subsec-Matrices-Vectors}Matrices as Vectors and Vectors as Matrices}
\pst Let us consider the linear space of $m\times n$ matrices and
choose a set of $mn$ matrix units $E_{(i,j)}$, $1 \le i \le m$, $1
\le j \le n$, as basis in this space:
\begin{equation} \label{matrix-unit}
\underset{m\times n}{E_{(i,j)}} = \bordermatrix{
  &   &   & \underset{\downarrow}{j} &  \cr
  & 0 & 0 & 0 & 0 & 0 \cr
  & 0 & \cdots & 0 & \cdots & 0 \cr
i \rightarrow & 0 & 0 & 1 & 0 & 0 \cr
  & 0 & \cdots & 0 & \cdots & 0 \cr
  & 0 & 0 & 0 & 0 & 0 \cr }.
\end{equation}

We use a known map (see, e.g.,~\cite{mmsz-03}) of $m\times n$
matrix $Z$ onto an $mn$-dimensional vector $|Z\rangle$ and vice
versa. For successive $i=1,2,\ldots,m$ take the $i$th row and
transpose it. Then join all the obtained $n$-columns step by step
to achieve the $mn$-dimensional column. This column is nothing
else but the coordinate representation of vector $|Z\rangle$ in
some orthogonal basis. For instance, in case $m=n=2$ we have
\begin{equation}
\label{matrices-vectors}
Z = \left(%
\begin{array}{cc}
  a & b \\
  c & d \\
\end{array}%
\right) \longrightarrow |Z\rangle = \left(%
\begin{array}{c}
  a \\
  b \\
  c \\
  d \\
\end{array}%
\right).
\end{equation}

Thus, thanks to this rule any rectangular $m\times n$ matrix can
be considered as $mn$-dimensional vector. Apparently, there exists
an inverse operation which provides the inverse map of a
$N$-dimensional vector $|Z\rangle$ onto matrix $Z$ if the number
of vector elements is a composite number $N=mn$. Such a composite
number $N=mn$ provides two rectangular matrices of dimension
$m\times n$ and $n\times m$, with matrices depending on how we
split up the vector onto components and then collect them in
columns and rows. This map provides the possibility to consider
any composite column vector as a $m\times n$ matrix of an
operator: $\mathcal{H}_n \rightarrow \mathcal{H}_m$. Conversely,
another matrix (of dimension $n\times m$) yields the map of a
vector from $\mathcal{H}_m$ onto a vector in $\mathcal{H}_n$.

The feature of a prime number $N$ is that the $N$-dimensional
vector cannot be bijectively mapped (without extension) onto a
$m\times n$ matrix with $m,n>1$. This characteristic property of
prime numbers can shad some light on proving nonexistence of a
full set of mutually unbiased bases in Hilbert spaces of
non-power-prime dimensions.

\textbf{Remark 1}. Square matrix $Z$ ($m=n=d$) is represented by
$d^2$-vector with components ${\rm Tr}\big[ E_{(i,j)}^{\dag} Z
\big]$. However, instead of matrix units $E_{(i,j)}$, one can use
another orthonormal (in trace sense) basis of matrices in
$\mathcal{B}(\mathcal{H}_d)$. For example, if $d=2$ one can use
conventional matrices of operators $\frac{1}{\sqrt{2}} (\hat{I}_2,
\hat{\sigma}_x, \hat{\sigma}_y, \hat{\sigma}_z)$, where
$\hat{I}_2\in\mathcal{B}(\mathcal{H}_2)$ is the identity operator
and $(\hat{\sigma}_x,\hat{\sigma}_y,\hat{\sigma}_z)$ is the set of
Pauli operators. Then
\begin{equation}
\label{matrices-vectors-Pauli}
Z = \left(%
\begin{array}{cc}
  a & b \\
  c & d \\
\end{array}%
\right) \longrightarrow |\widetilde{Z}\rangle = \frac{1}{\sqrt{2}}\left(%
\begin{array}{c}
  a+d \\
  b+c \\
  i(b-c) \\
  a-d \\
\end{array}%
\right).
\end{equation}

\subsection{\label{subsec-Hierarchy-Operators}Hierarchy of Operators}
\pst Applying the above consideration to $d\times d$ matrices $X$
and $Y$ of operators
$\hat{X},\hat{Y}\in\mathcal{B}(\mathcal{H}_d)$ results in
$d^2$-dimensional complex vectors $|X\rangle$ and $|Y\rangle$ such
that $\langle X | Y \rangle = {\rm Tr}\big[ \hat{X}^{\dag} \hat{Y}
\big]$. In other words, trace operation applied to the product of
two matrices is equivalent to the standard scalar product of
column vectors constructed from the initial matrices. It follows
easily that $\mathcal{B}(\mathcal{H}_d)$ is isomorphic to
$\mathcal{H}_{d^2}$, i.e. $\mathcal{B}(\mathcal{H}_d)
\Longleftrightarrow \mathcal{H}_{d^2}$. On obtaining this crucial
result one can readily repeat the development of this Section by
substituting $d^2$ for $d$. Similarly, one can construct a
$d^4$-dimensional Hilbert space
$\mathcal{B}(\mathcal{B}(\mathcal{H}_d))$ of operators acting on
the space of operators $\mathcal{B}(\mathcal{H}_d)$ which in turn
act on vectors from $\mathcal{H}_d$. We will refer to the space
$\mathcal{B}(\mathcal{B}(\mathcal{H}_d))$ as a space of
superoperators on $\mathcal{H}_d$. Evidently,
$\mathcal{B}(\mathcal{B}(\mathcal{H}_d)) \Longleftrightarrow
\mathcal{H}_{d^4}$ and this consideration can be continued ad
infinitum. This leads to the following hierarchy of spaces:
\begin{equation}
\mathcal{H}_d \Longrightarrow \mathcal{B}(\mathcal{H}_d)
\Longleftrightarrow \mathcal{H}_{d^2} \Longrightarrow
\mathcal{B}(\mathcal{B}(\mathcal{H}_d)) \Longleftrightarrow
\mathcal{B}(\mathcal{H}_{d^2}) \Longleftrightarrow
\mathcal{H}_{d^4} \Longrightarrow \ldots
\end{equation}

\section{\label{section-generic-star-product-scheme} Star Product for Discrete Variables}
\pst In this Section, following the ideas
of~\cite{oman'ko-JPA,oman'ko-vitale,ibort} we review a
construction of the star product for functions depending on
discrete variables.

Let us consider the Hilbert space $\mathcal{B}(\mathcal{H}_d)$.

{\bf Definition}. The function $f_{A}(k)$ on a discrete set
$\{k\}$, $k=1,\ldots,N<\infty$, defined by the relation
\begin{equation}
\label{dequantizer} f_A(k) = {\rm Tr} \big[\hat{U}_k^{\dag}
\hat{A}\big]
\end{equation}
\noindent is called the symbol of an operator
$\hat{A}\in\mathcal{B}(\mathcal{H}_d)$ and an operator
$\hat{U}_k\in\mathcal{B}(\mathcal{H}_d)$ is called dequantizer
operator of the star-product scheme.

Note that a symbol $f_A(k)$ can be considered as elements of a
column $\boldsymbol{f}_A = \left(%
\begin{array}{ccc}
  f_A(1) & \cdots & f_A(N) \\
\end{array}%
\right)^{\rm tr}$. For example, if we choose $d\times d$ matrix
units $\hat{E}_{(i,j)}$ as quantizers $\hat{U}_k$, where the index
$k=1,\ldots,d^2$ is parameterized by $k=d(i-1)+j$, then
$\boldsymbol{f}_A = |A\rangle \in \mathcal{H}_{d^2}$.

If the symbol $f_A(k)$ contains a full information about the
operator $\hat{A}$, then such star-product scheme is tomographic
(informationally complete). In other words, knowledge of the
symbol $f_A(k)$ is sufficient in order to find an explicit form of
the operator $\hat{A}$, namely,
\begin{equation}
\label{quantizer} \hat{A} = \sum_{k=1}^{N} f_A(k) \hat{D}_k.
\end{equation}
\noindent The operator $\hat{D}_k \in \mathcal{B}(\mathcal{H}_d)$
is referred to as quantizer and is connected with the dequantizer
$\hat{U}_{k'}$ by means of relation
\begin{equation}
\label{delta} {\rm Tr} \big[ \hat{U}_{k}^{\dag} \hat{D}_{k'} \big]
= \delta(k,k'),
\end{equation}
\noindent where the function $\delta(k,k')$ of two discrete
variables plays a role of delta-function on the set of tomographic
symbols of all operators. In other words,
\begin{equation}
\label{delta-check} \sum_{k'=1}^{N} f_A(k') \delta(k,k') = f_A(k).
\end{equation}

\subsection{\label{subsec-Tomographic-Star-Product-Scheme} Tomographic Star-Product Scheme}
\pst It is shown
in~\cite{manko-marmo-simoni-etal,manko-marmo-simoni-vent,mms-sudarshan-vent}
that the star-product scheme (\ref{dequantizer}),
(\ref{quantizer}) is tomographic if and only if
\begin{equation}
\label{tomogr-s-p-require} \sum_{k=1}^{N} | D_k \rangle \langle
U_k | = \hat{I}_{d^2},
\end{equation}
\noindent where $|D_k \rangle,|U_k \rangle\in\mathcal{H}_{d^2}$
are vectors constructed from the quantizer $\hat{D}_k$ and the
dequantizer $\hat{U}_k$, respectively, by the higher-dimensional
analog of the rule (\ref{matrices-vectors}), $\langle U_k | = |U_k
\rangle^{\dag}$, and $\hat{I}_{d^2}$ is an identity operator in
$\mathcal{B}(\mathcal{H}_{d^2})$. It is worth noting that
condition (\ref{delta-check}) is then automatically met because
$\delta(k,k') = \langle U_k | D_{k'} \rangle$, $f_A(k') = \langle
U_{k'} | A \rangle$, and $\sum_{k'=1}^{N} \langle U_k | D_{k'}
\rangle \langle U_{k'} | A \rangle = \langle U_k | \hat{I}_{d^2} |
A \rangle = \langle U_k | A \rangle$.

An evident requirement for (\ref{tomogr-s-p-require}) to be
fulfilled is $N\ge d^2$, because a sum of rank-1 projectors should
be equal to the full-rank operator. For the inverse map
(\ref{quantizer}): $\mathbb{C}^{N} \rightarrow
\mathcal{B}(\mathcal{H}_d)$ to exist, it is necessary and
sufficient that the set of dequantizers $\{\hat{U}_k\}_{k=1}^{N}$
contains $d^2$ linearly independent operators. If we combine the
corresponding $d^2$-dimensional columns $|U_k\rangle$ into a
single $d^2\times N$ dequantization matrix $\mathscr{U}$ of the
form
\begin{equation}
\label{dequantization-matrix} \underset{d^2\times N}{\mathscr{U}}
= \left( \Bigg| U_1 \Bigg\rangle \Bigg| U_2 \Bigg\rangle \cdots
\Bigg| U_N \Bigg\rangle \right)  = \left(
\begin{array}{cccc}
  |U_1\rangle_1 & |U_2\rangle_1 & \cdots & |U_N\rangle_1\\
  |U_1\rangle_2 & |U_2\rangle_2 & \cdots & |U_N\rangle_2\\
  \cdots & \cdots & \cdots & \cdots\\
  |U_1\rangle_{d^2} & |U_2\rangle_{d^2} & \cdots & |U_N\rangle_{d^2}\\
\end{array} \right),
\end{equation}
\noindent then this criterion can be rewritten as ${\rm rank}
\mathscr{U} = d^2$. Once this condition is met, a set of
quantizers $\{\hat{D}_k\}_{k=1}^{N}$ exists and can also be
written in terms of a single quantization matrix
\begin{equation}
\label{quantization-matrix} \underset{d^2\times N}{\mathscr{D}} =
\left( \Bigg| D_1 \Bigg\rangle \Bigg| D_2 \Bigg\rangle \cdots
\Bigg| D_N \Bigg\rangle \right)  = \left(
\begin{array}{cccc}
  |D_1\rangle_1 & |D_2\rangle_1 & \cdots & |D_N\rangle_1\\
  |D_1\rangle_2 & |D_2\rangle_2 & \cdots & |D_N\rangle_2\\
  \cdots & \cdots & \cdots & \cdots\\
  |D_1\rangle_{d^2} & |D_2\rangle_{d^2} & \cdots & |D_N\rangle_{d^2}\\
\end{array} \right).
\end{equation}

In Section \ref{sec-Dequantization-Matrix-and-SPScheme}, we will
reveal a relation between matrices $\mathscr{U}$, $\mathscr{D}$
and properties of the star-product scheme.

\textbf{Remark 2}. Exploiting the notation
(\ref{dequantization-matrix})--(\ref{quantization-matrix}), the
criterion (\ref{tomogr-s-p-require}) takes the form
$\mathscr{D}\mathscr{U}^{\dag} = I_{d^2}$.

\subsubsection{\label{subsubsec-Search-Quantization-Matrix}Search of Quantization Matrix}
\pst Given the dequantization matrix $\mathscr{U}$, ${\rm
rank}\mathscr{U} = d^2$, a quantization matrix
(\ref{quantization-matrix}) can be found via the following
pseudoinverse operation
\begin{equation}
\label{D-from-U} \mathscr{D} =
(\mathscr{U}\mathscr{U}^{\dag})^{-1} \mathscr{U}.
\end{equation}
\noindent Indeed, it can be easily checked that
$\sum_{k=1}^{N}|U_k\rangle\langle U_k| =
\mathscr{U}\mathscr{U}^{\dag}$. Hence,
\begin{equation}
\sum_{k=1}^{N} | D_k \rangle \langle U_k | = \sum_{k=1}^{N}
(\mathscr{U}\mathscr{U}^{\dag})^{-1} |U_k\rangle\langle U_k| =
(\mathscr{U}\mathscr{U}^{\dag})^{-1} \mathscr{U}\mathscr{U}^{\dag}
= \hat{I}_{d^2},
\end{equation}
\noindent i.e. the requirement (\ref{tomogr-s-p-require}) holds
true.

It is worth mentioning that the matrix $\mathcal{D}$ does not have
to be expressed in the form (\ref{D-from-U}) if $N>d^2$. In fact,
in this case vectors $\{|U_k\rangle\}_{k=1}^{N}$ are linearly
dependent. Therefore there exists a nontrivial linear combination
$\sum_{k=1}^{N}c_k |U_k\rangle = 0$. Transformation $\delta(k,k')
\rightarrow \delta(k,k')+c_{k'}^{\ast}$ leaves the equality
(\ref{delta-check}) accomplished. Such a transformation is easily
achieved by the following transformation of the quantization
matrix: $\mathscr{D} \rightarrow \mathscr{D} + (\cdot){\rm
diag}(c_1^{\ast}, c_2^{\ast}, \ldots, c_N^{\ast})$, where
$(\cdot)$ is an arbitrary $d^2\times N$ matrix. This means that an
ambiguity of quantization matrix (\ref{quantization-matrix}) is
allowed and formula (\ref{D-from-U}) covers only one of many
possibilities.

\subsubsection{\label{subsubsec-MinimalTSPS} Minimal Tomographic Star-Product Scheme}
\pst Important is the special case $N=d^2$ leading to a
\textit{minimal} tomographic star-product scheme. The condition
${\rm rank}\mathscr{U} = d^2$ is then equivalent to
$\det\mathscr{U} \ne 0$, i.e. to the existence of the inverse
matrix $\mathscr{U}^{-1}$. Formula (\ref{delta-check}) is valid
for any symbol $f_A(k)$, $k=1,\ldots,d^2$ if and only if
$\delta(k,k')$ reduces to the Kronecker delta-symbol
$\delta_{k,k'}$. Taking into account relation (\ref{delta}), we
obtain
\begin{equation}
\label{D-from-U-square-matrix}
\mathscr{U}^{\dag}\mathscr{D}=I_{d^2} \quad \Longleftrightarrow
\quad \mathscr{D} = (\mathscr{U}^{\dag})^{-1}.
\end{equation}

\textbf{Example 1}. It is easily seen that if we choose $d\times
d$ matrix units $\hat{E}_{(i,j)}$ as dequantizers $\hat{U}_k$,
$k=d(i-1)+j$, then $\mathscr{U}=\mathscr{D}=I_{d^2}$ and the
requirement (\ref{tomogr-s-p-require}) is satisfied. Such a
tomographic procedure results in the proper reconstruction formula
(\ref{quantizer}) with $\hat{D}_k = \hat{E}_{(i,j)}$. However, in
physics, scientists are interested in the reconstruction of the
Hermitian density operator $\hat{\rho}$ by measuring physical
quantities associated with Hermitian dequantizer operators
$\hat{U}_{k}=\hat{U}_{k}^{\dag}$ (in contrast to matrix units for
which $\hat{E}_{(i,j)}^{\dag} = \hat{E}_{(j,i)} \ne
\hat{E}_{(i,j)}$). The most general case of measurements
associated with positive operator-valued measures is considered in
Section
\ref{subsec-square-matrix}.\hfill$\scriptstyle\blacksquare$

\subsection{\label{subsec-Star-Product-Kernel} Star-Product
Kernel} \pst The symbol $f_{AB}(k)$ of the product of two
operators $\hat{A},\hat{B}\in\mathcal{B}(\mathcal{H}_d)$ equals a
star product of symbols $f_A$ and $f_B$ determined by the formula
\begin{equation}
(f_A \star f_B) (k) \equiv f_{AB}(k) = \sum_{k',k''=1}^{N} f_A(k')
f_B(k'') K(k,k',k''),
\end{equation}
\noindent where the kernel $K$ is expressed in terms of
dequantizer and quantizer operators as follows:
\begin{equation}
K(k,k',k'') = {\rm Tr} \big[ \hat{U}_{k}^{\dag} \hat{D}_{k'}
\hat{D}_{k''} \big].
\end{equation}
\noindent Since star product is associative by definition, it
necessarily satisfies the nonlinear equation
\begin{equation}
K^{(3)}(k,k',k'',k''') = \sum_{l=1}^{N} K(k,l,k''') K(l,k',k'') =
\sum_{l=1}^{N} K(k,k',l) K(l,k'',k'''),
\end{equation}
\noindent which is an immediate consequence of the relation $f_A
\star f_B \star f_C = (f_A \star f_B) \star f_C = f_A \star (f_B
\star f_C)$.

\subsection{\label{subsec-Interteining-Kernels} Intertwining Kernels
Between Two Star-Product Schemes}
\pst Let us assume that we are
given two different discrete sets $\{k\}_{k=1}^{N}$ and
$\{\kappa\}_{\kappa=1}^{M}$ as well as two different sets of the
corresponding dequantizers and quantizers,
$\{\hat{U}_k,\hat{D}_k\}_{k=1}^{N}$ and
$\{\hat{\mathfrak{U}}_{\kappa},\hat{\mathfrak{D}}_{\kappa}\}_{\kappa=1}^{M}$,
respectively, with operators from both sets acting on the same
Hilbert space $\mathcal{H}_d$. In view of this, one can construct
two different star-product schemes for two different kinds of
symbols $f_A(k)$ and $\mathfrak{f}_A(\kappa)$. The symbols are
related by intertwining kernels
\begin{eqnarray}
\label{intertw-symbols} f_A(k) = \sum_{\kappa=1}^{M}
K_{\mathfrak{f} \rightarrow f} (k,\kappa)
\mathfrak{f}_A(\kappa), \qquad \boldsymbol{f}_A = K_{\mathfrak{f} \rightarrow f} ~ \boldsymbol{\mathfrak{f}}_A, \nonumber\\
\mathfrak{f}_A(\kappa) = \sum_{k=1}^{N} K_{f \rightarrow
\mathfrak{f}} (\kappa,k) f_A(k), \qquad
\boldsymbol{\mathfrak{f}}_A = K_{f \rightarrow \mathfrak{f}} ~
\boldsymbol{f}_A,
\end{eqnarray}
\noindent where the intertwining kernels are represented as
rectangular matrices expressed through dequantizers and quantizers
as follows:
\begin{eqnarray}
\label{intertw-kernels} && K_{\mathfrak{f} \rightarrow f}
(k,\kappa) = {\rm Tr} \big[ \hat{U}_k^{\dag}
\hat{\mathfrak{D}}_{\kappa}\big], \qquad K_{f \rightarrow
\mathfrak{f}} (\kappa,k) = {\rm Tr} \big[
\hat{\mathfrak{U}}_{\kappa}^{\dag} \hat{D}_{k}\big],\\
&& K_{\mathfrak{f} \rightarrow f} = \mathscr{U}_{\{k\}}^{\dag} \mathscr{D}_{\{\kappa\}} = \underset{N\times M}{\left(%
\begin{array}{cc}
  \vdots & \vdots \\
\end{array}%
\right)}, \qquad K_{f
\rightarrow \mathfrak{f}} = \mathscr{U}_{\{\kappa\}}^{\dag} \mathscr{D}_{\{k\}} = \underset{M\times N}{\left(%
\begin{array}{c}
  \cdots\\
  \cdots\\
\end{array}%
\right)}.
\end{eqnarray}

\textbf{Example 2}. Given a unitary $d^2\times d^2$ matrix $u$, we
construct two star-product schemes: the first one exploits columns
of the matrix $u$ as dequantizers $|U_k\rangle$ (i.e.
$\mathscr{U}_{\{k\}}=\mathscr{D}_{\{k\}}=u$), the second one
utilizes rows of the matrix $u$ as dequantizers
$|\mathfrak{U}_k\rangle$ (i.e.
$\mathscr{U}_{\{\kappa\}}=\mathscr{D}_{\{\kappa\}}=u^{\rm tr}$).
Using formulas (\ref{intertw-symbols}), (\ref{intertw-kernels})
and decomposing row matrix elements in terms of column matrix
elements, we get the cubic relation $u = (uu^{\ast})u^{\rm tr}$.
\hfill$\scriptstyle\blacksquare$

One can consider a particular case $\{k\} \equiv \{\kappa\}$,
$\hat{\mathfrak{U}}_{\kappa} = \hat{D}_{k}$, and
$\hat{\mathfrak{D}}_{\kappa} = \hat{U}_{k}$, which is called dual
star-product quantization scheme.

\subsection{\label{subsec-Self-Dual} Self-Dual Star-Product Scheme}
\pst \textbf{Definition}. Star-product scheme (\ref{dequantizer}),
(\ref{quantizer}) is called self-dual if there exists
$c\in\mathbb{R}$, $c>0$ such that $\hat{U}_k = c \hat{D}_k$ for
all $k=1,\ldots,N$. We will refer to the factor $c$ as coefficient
of skewness.

Self-dual star-product scheme is completely equivalent to the
scheme with coincident dequantizer and quantizer operators
$\hat{\widetilde{U}}_k = \hat{\widetilde{D}}_k =
\frac{1}{\sqrt{c}} \hat{U}_k = \sqrt{c} \hat{D}_k$.

\textbf{Example 3}. Matrix units $\hat{E}_{(i,j)}$ form a
self-dual scheme with $c=1$.\hfill$\scriptstyle\blacksquare$

\textbf{Example 4}. A description of the qubit ($d=2$) phase space
proposed in the paper~\cite{livine} implies a self-dual
star-product scheme with the following dequantizers and
quantizers:
\begin{eqnarray}
\label{qubit-phase} && \hat{U}_1 = \frac{1}{2}\hat{D}_1 =
\frac{1}{4}\left( \hat{I}_2 + \hat{\sigma}_x + \hat{\sigma}_y +
\hat{\sigma}_z \right),
\nonumber\\
&& \hat{U}_2 = \frac{1}{2}\hat{D}_2 = \frac{1}{4}\left( \hat{I}_2
+ \hat{\sigma}_x - \hat{\sigma}_y - \hat{\sigma}_z \right),
\nonumber\\
&& \hat{U}_3 = \frac{1}{2}\hat{D}_3 = \frac{1}{4}\left( \hat{I}_2
- \hat{\sigma}_x + \hat{\sigma}_y - \hat{\sigma}_z \right),
\nonumber\\
&& \hat{U}_4 = \frac{1}{2}\hat{D}_4 = \frac{1}{4}\left( \hat{I}_2
- \hat{\sigma}_x - \hat{\sigma}_y + \hat{\sigma}_z \right).
\end{eqnarray}\hfill$\scriptstyle\blacksquare$

\section{\label{sec-Dequantization-Matrix-and-SPScheme} Type of Dequantization Matrix and Properties of Star-Product Scheme}
\pst In this Section, we will establish a relation between the
type of dequantization matrix $\mathscr{U}$ (quantization matrix
$D$) and particular properties of the star-product scheme. Unless
specifically stated, we deal with the $d^2$-dimensional space of
operators $\mathcal{B}(\mathcal{H}_d)$.

\subsection{Rectangular Matrix}
We start with the most general rectangular $d^2\times N$ matrix
$\mathscr{U}$. As it was shown previously in Section
\ref{subsec-Tomographic-Star-Product-Scheme}, if $N<d^2$ then
${\rm rank}\mathscr{U} \le N < d^2$, the set of dequantizers
$\{\hat{U}_k\}_{k=1}^{N}$ is underfilled, and quantization matrix
$D$ is not defined. In the opposite case $N\ge d^2$, the scheme is
underfilled again if ${\rm rank}\mathscr{U} < d^2$ and the scheme
is overfilled if ${\rm rank}\mathscr{U} = d^2$. Underfilled
schemes enable revealing partial information about the system. The
greater ${\rm rank}\mathscr{U}$ the more information can can be
extracted from the symbols (\ref{dequantizer}). Under this
circumstance, the closer $N$ to ${\rm rank}\mathscr{U}$, the less
resource-intensive is the procedure. Overfilled set of dequntizers
provides a tomographic star-product scheme and allows calculating
quantization matrix $\mathscr{D}$, e.g. according to formula
(\ref{D-from-U}). For overfilled scheme, the smaller difference
$N-d^2$ the less redundant information is contained in tomographic
symbols.

\textbf{Example 5}. Consider a full set of mutually unbiased bases
(MUBs) $\{|a\alpha\rangle\}$, $a=0,\ldots,d$ (basis number),
$\alpha=0,\ldots,d-1$ (vector index inside a basis) in
power-prime-dimensional Hilbert space $\mathcal{H}_d$.
Dequantizers of the form $|a\alpha\rangle \langle a\alpha | \in
\mathcal{B}(\mathcal{H}_d)$ lead to an overfilled scheme with the
$d^2\times d(d+1)$ rectangular dequantization matrix
$\mathscr{U}$, ${\rm rank}\mathscr{U} = d^2$. The case $d=2$ is
illustrated in Table \ref{table}.\hfill$\scriptstyle\blacksquare$

\subsection{\label{subsec-square-matrix}Square Matrix}
\pst An arbitrary square $d^2\times d^2$ matrix $\mathscr{U}$ with
$\det \mathscr{U} \ne 0$ defines a minimal tomographic
star-product scheme and vice versa. Quantization matrix
$\mathscr{D}$ is given by formula (\ref{D-from-U-square-matrix}).
Symbols (\ref{dequantizer}) thoroughly  determine a desired
operator $\hat{A}\in\mathcal{B}(\mathcal{H}_d)$. The density
operator $\hat{\rho}$ of the physical system is of special
interest. All informationally complete positive operator-valued
measures (POVMs) are nothing else but either overfilled or minimal
tomographic star-product schemes (see, e.g.,~\cite{weigert}),
where POVM effects are regarded as dequantizers. If this is the
case, symbols can, in principal, be measured experimentally.
Assuming a non-zero error bar of measured symbols, the less is the
condition number of the matrix $\mathscr{U}$ the less erroneous is
the reconstructed density operator (in a desired basis).

\textbf{Example 6}. Symmetric informationally complete POVM
(SIC-POVM) of the Weyl-Heisenberg form is conjectured to exist for
an arbitrary finite dimension $d=\dim\mathcal{H}_d$ (although not
proven yet). SIC-POVM consists of $d^2$ effects $\hat{U}_k =
\frac{1}{d} \hat{\Pi}_k =
\frac{1}{d}|\psi_k\rangle\langle\psi_k|\in\mathcal{B}(\mathcal{H}_d)$
such that ${\rm Tr}\big[ \hat{\Pi}_k \hat{\Pi}_{k'} \big] =
(d\delta_{kk'}+1)/(d+1)$. It means that the scalar product
$\langle U_k | U_{k'} \rangle$ of any two different columns of
matrix $\mathscr{U}$ is the same number $1/d^2(d+1)$. The example
of qubits is placed in the Table \ref{table}.
\hfill$\scriptstyle\blacksquare$

\begin{figure}
\begin{center}
\includegraphics{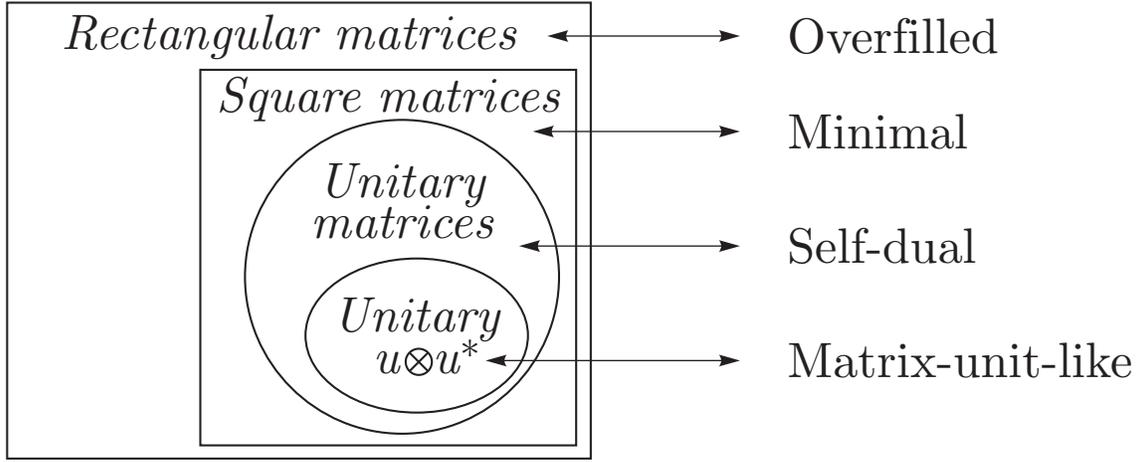}
\caption{\label{figure} One-to-one correspondence between the type
of dequantization matrix $\mathscr{U}$, ${\rm rank}\mathscr{U} =
d^2$, and the type of star-product scheme in $\mathcal{H}_d$. The
matrix $\mathscr{U}$ is constructed by higher-dimensional
analogues of formulas (\ref{matrices-vectors}),
(\ref{dequantization-matrix}).}
\end{center}
\end{figure}

\subsection{Unitary Matrix}
\pst To begin with, let us remind some properties of unitary
matrices. A unitary $d^2\times d^2$ matrix $\mathscr{U}$ satisfies
the condition $\mathscr{U} \mathscr{U}^{\dag} = \mathscr{U}^{\dag}
\mathscr{U} = I_{d^2}$. This property implies the orthogonality of
columns of this matrix
\begin{equation}
\sum_{p=1}^{d}\mathscr{U}_{pq}^{\ast} \mathscr{U}_{pq'} =
\delta_{qq'},
\end{equation}
\noindent It can be easily checked that the rows are also
orthogonal, i.e. $\sum_{q=1}^{d} \mathscr{U}_{pq}^{\ast}
\mathscr{U}_{p'q} = \delta_{pp'}$. This property means that the
columns (rows) of the matrix $\mathscr{U}$ can be chosen as
orthonormal basis vectors in $d^2$-dimensional Hilbert space
$\mathcal{H}_{d^2}$ and, consequently, in the space
$\mathcal{B}(\mathcal{H}_d)$ by the higher-dimensional analogue of
the map inverse to (\ref{matrices-vectors}). It means that all
bases and sets of operators in $\mathcal{B}(\mathcal{H}_{d})$ can
be represented as linear combinations of operators ${\hat{U}_k}$
obtained from the columns $|U_k\rangle$ of matrix $\mathscr{U}$.

Now, we proceed to the analysis of the relation between the
unitary dequantization matrix $\mathscr{U}$ and features of the
star-product scheme.

\textbf{Proposition 1}. A star-product scheme is minimal self-dual
with coefficient of skewness $c$ if and only if the corresponding
dequantization matrix $\mathscr{U}=\sqrt{c}\tilde{\mathscr{U}}$,
where $\tilde{\mathscr{U}}$ is a unitary $d^2\times d^2$ matrix.

\textbf{Proof}. As it is stated in Section \ref{subsec-Self-Dual},
a self-dual star-product scheme is equivalent to the scheme with
coincident quantizers and dequantizers, i.e. $\tilde{\mathscr{U}}
= \tilde{\mathscr{D}} = \frac{1}{\sqrt{c}}\mathscr{U}$. On the
other hand, from (\ref{D-from-U-square-matrix}) it follows that
$\tilde{\mathscr{U}}^{\dag} = \tilde{\mathscr{U}}^{-1}$. Now the
statement of the Proposition is clearly seen.
\hfill$\scriptstyle\blacksquare$

For many applications it is important to be aware of the relation
between POVMs (primarily used for performing tomography of the
system) and self-dual schemes (usually exploited while considering
phase-space of the system). The following Propositions reveal an
incompatibility of these two approaches.

\textbf{Proposition 2}. There exists no minimal tomographic
star-product scheme with dequantizers in the form of POVM effects
and Hermitian semi-positive quantizers.

\textbf{Proof}. Assume the converse, namely,
$\sum_{k=1}^{d^2}\hat{U}_k = \hat{I}_{d}$, $\hat{U}_k =
\hat{U}_k^{\dag} \ge 0$, and $\hat{D}_k = \hat{D}_k^{\dag} \ge 0$
for all $k=1,\ldots,d^2$. From Eq. (\ref{D-from-U-square-matrix})
it follows that ${\rm Tr}\big[ \hat{U}_k \hat{D}_{k'} \big] =
\delta_{kk'}$ and $\sum_{k=1}^{d^2}{\rm Tr}\big[ \hat{U}_k
\hat{D}_{k'} \big] = {\rm Tr}\big[\hat{D}_{k'} \big] = 1$. This
implies that $\{\hat{D}_{k}\}_{k=1}^{d^2}$ is a set of density
operators. Since $0\le \hat{U}_k \le \hat{I}_{d}$ then the
equality ${\rm Tr}\big[ \hat{U}_k \hat{D}_{k} \big] = 1$ can be
only achieved if $\hat{U}_k=\hat{D}_{k}=| \psi_k \rangle\langle
\psi_k |$, $| \psi_k \rangle\in\mathcal{H}_d$ or $\hat{U}_k =
\hat{I}_d$. The latter case is inconsistent in view of POVM
requirement $\sum_{k=1}^{d^2}\hat{U}_k = \hat{I}_{d}$ and the
former case implies $\langle \psi_k | \psi_{k'} \rangle =
\delta_{kk'}$ for all $k,k'=1,\ldots,d^2$, which is impossible as
there can be no greater than $d$ orthonormal vectors in
$\mathcal{H}_d$. This contradiction concludes the proof.
\hfill$\scriptstyle\blacksquare$

This proposition is followed by immediate consequences.

\textbf{Corollary 1}. There exists no minimal self-dual
star-product scheme with dequantizers in the form of POVM effects.

\textbf{Proof}. If such a scheme existed, then the quantizers
would be Hermitian semi-positive in view of self duality. This
contradicts to Proposition 1. \hfill$\scriptstyle\blacksquare$

\textbf{Corollary 2}. If dequantizers $\{\hat{U}_k\}_{k=1}^{d^2}$
form a POVM, then dequantization and quantization matrices
$\mathscr{U}$ and $\mathscr{D}$ are not proportional to any
unitary matrix.

\textbf{Corollary 3}. Hermitian dequantizers and quantizers of a
self-dual scheme must contain negative eigenvalues.

The result of Corollary 1 indicates a slight error in the
paper~\cite{livine}, where dequantizers of the self-dual scheme
(\ref{qubit-phase}) are treated as POVM effects, which is
incorrect but harmless to the rest of the article. The
paper~\cite{appleby-arbitrary-rank} uses a notation ``Wigner POVM"
because of an observed connection of Wigner function with
POVM-probabilities rescaled by a constant amount and then shifted
by a constant amount. The very shift makes the scheme
non-self-dual (as it should be according to Corollary 1). Taking
into account Proposition 2, we can predict the negative sign of
this shift.

The obtained results seem to be valid not only in
finite-dimensional Hilbert spaces but also in infinite dimensional
case. For instance, Corollary 3 is illustrated by the following
example.

\textbf{Example 7}. Weyl star-product scheme is defined through
dequantizers $\hat{U}(q,p) = 2 \hat{\mathcal{D}}(\alpha)
\hat{\mathcal{I}} \hat{\mathcal{D}}(-\alpha)$ and quantizers
$\hat{D}(q,p) = \frac{1}{2\pi}\hat{U}(q,p)$, where
$\alpha=(q+ip)/\sqrt{2}$, $\hat{\mathcal{D}}(\alpha) = \exp\big[
\alpha \hat{a}^{\dag} - \alpha^{\ast} \hat{a} \big]$ is the
displacement operator, $\hat{a}^{\dag}$ and $\hat{a}$ are creation
and annihilation operators, respectively, $\hat{\mathcal{I}}$ is
the inversion operator. The scheme is obviously self-dual. Since
the displacement operator is unitary, dequantizers and quantizers
are Hermitian and inherit a spectrum of the inversion operator
${\rm Sp}_{\mathcal{I}}=\{\pm 1\}$, i.e. exhibit negative
eigenvalues. \hfill$\scriptstyle\blacksquare$

\subsection{Unitary Matrix $u\otimes u^{\ast}$}
\pst The dequantization matrix of the form $u\otimes u^{\ast}$
occurs while performing a unitary rotation of matrix units
$\hat{E}_{(i,j)}$, $i,j=1,\ldots,d$. Indeed, a transform $u
E_{(i,j)} u^{\dag} = | u_i \rangle \langle u_j |$, where $| u_i
\rangle$ is the $i$th column of a unitary $d\times d$ matrix $u$,
$\langle u_j | = | u_j \rangle^{\dag}$. Vector representation
(\ref{matrices-vectors}) of the matrix $| u_i \rangle \langle u_j
|$ is $| u_i \rangle \otimes (\langle u_j |)^{\rm tr} = | u_i
\rangle \otimes (| u_j \rangle)^{\ast}$. Stacking these vectors by
the rule (\ref{dequantization-matrix}) yields
$\mathscr{U}=u\otimes u^{\ast}$. It means that such a matrix
$\mathscr{U}$ defines dequantizers and quantizers of the form
$\hat{U}_k = \hat{D}_k = \hat{u} \hat{E}_{(i,j)} \hat{u}^{\dag} =
\hat{u} | e_i \rangle \langle e_j | \hat{u}^{\dag} = | \psi_i
\rangle \langle \psi_j |$ for all $k=1,\ldots,d^2$. It is worth
noting that $\langle \psi_i | \psi_j \rangle = \delta_{ij}$, so
the star-product scheme is matrix-unit-like, with all dequantizers
and quantizers being rank-1 operators.

The results of this Section concerning tomographic star-product
schemes are depicted in Figure \ref{figure}. We also provide a
summary Table \ref{table} of examples for qubits.

\begin{table}[t]
\caption{\label{table} Examples of $4 \times N$ matrices
$\mathscr{U}$ and corresponding bases (sets of vectors) in
$\mathcal{H}_2$}
\begin{tabular}{|c|c|c|}
  \hline
 Dequantizers &  Dequantization matrix $\mathscr{U}$ & Dequantization matrix $\mathscr{U}$\\
 $\{\hat{U}_k\}_{k=1}^{N}$ & constructed by rules (\ref{matrices-vectors}), (\ref{dequantization-matrix}) & constructed by rules (\ref{matrices-vectors-Pauli}), (\ref{dequantization-matrix}) \\

  \hline
  $\begin{array}{c}
  {\rm Matrix~units} \\
  \hat{E}_{(i,j)},~ {\rm Eq.}~(\ref{matrix-unit}) \\
\end{array}$ & $\left(%
\begin{array}{cccc}
  1 & 0 & 0 & 0 \\
  0 & 1 & 0 & 0 \\
  0 & 0 & 1 & 0 \\
  0 & 0 & 0 & 1 \\
\end{array}%
\right)$ & $\frac{1}{\sqrt{2}}\left(%
\begin{array}{cccc}
  1 & 0 & 0 & 1 \\
  0 & 1 & 1 & 0 \\
  0 & i & -i & 0 \\
  1 & 0 & 0 & -1 \\
\end{array}%
\right)$\\
  \hline
  $\frac{1}{\sqrt{2}}(\hat{I}_2, \hat{\sigma}_x, \hat{\sigma}_y, \hat{\sigma}_z)$ & $\frac{1}{\sqrt{2}}\left(%
\begin{array}{cccc}
  1 & 0 & 0 & 1 \\
  0 & 1 & -i & 0 \\
  0 & 1 & i & 0 \\
  1 & 0 & 0 & -1 \\
\end{array}%
\right)$ & $\left(%
\begin{array}{cccc}
  1 & 0 & 0 & 0 \\
  0 & 1 & 0 & 0 \\
  0 & 0 & 1 & 0 \\
  0 & 0 & 0 & 1 \\
\end{array}%
\right)$ \\
  \hline
$\frac{1}{\sqrt{2}}(\hat{I}_2, \hat{\sigma}_x, i\hat{\sigma}_y, \hat{\sigma}_z)$ & $\frac{1}{\sqrt{2}}\left(%
\begin{array}{cccc}
  1 & 0 & 0 & 1 \\
  0 & 1 & 1 & 0 \\
  0 & 1 & -1 & 0 \\
  1 & 0 & 0 & -1 \\
\end{array}%
\right)$ & $\left(%
\begin{array}{cccc}
  1 & 0 & 0 & 0 \\
  0 & 1 & 0 & 0 \\
  0 & 0 & i & 0 \\
  0 & 0 & 0 & 1 \\
\end{array}%
\right)$\\
  \hline
$\begin{array}{c}
  {\rm Eqs.~(\ref{qubit-phase})} \\
  {\rm (Ex.~4)} \\
\end{array}$ & $\frac{1}{2}\left(%
\begin{array}{cccc}
  2 & 0 & 0 & 2 \\
  1-i & 1+i & -1-i & -1+i \\
  1+i & 1-i & -1+i & -1-i \\
  0 & 2 & 2 & 0 \\
\end{array}%
\right)$ & $\frac{1}{2}\left(%
\begin{array}{cccc}
  1 & 1 & 1 & 1 \\
  1 & 1 & -1 & -1 \\
  1 & -1 & 1 & -1 \\
  1 & -1 & -1 & 1 \\
\end{array}%
\right)$\\
  \hline
$\begin{array}{c}
  {\rm SIC\!-\!POVM} \\
  {\rm (Ex.~6)} \\
\end{array}$ & $\frac{1}{2\sqrt{3}}\left(%
\begin{array}{cccc}
  \sqrt{3}+1 & \sqrt{3}-1 & \sqrt{3}-1 & \sqrt{3}+1 \\
  1-i & 1+i & -1-i & -1+i \\
  1+i & 1-i & -1+i & -1-i \\
  \sqrt{3}-1 & \sqrt{3}+1 & \sqrt{3}+1 & \sqrt{3}-1 \\
\end{array}%
\right)$ & $\frac{1}{2\sqrt{3}}\left(%
\begin{array}{cccc}
  \sqrt{3} & \sqrt{3} & \sqrt{3} & \sqrt{3} \\
  1 & 1 & -1 & -1 \\
  1 & -1 & 1 & -1 \\
  1 & -1 & -1 & 1 \\
\end{array}%
\right)$\\
  \hline
MUBs (Ex. 5) & $\frac{1}{\sqrt{2}}\left(%
\begin{array}{cccccc}
  \sqrt{2} & 0 & 1 & 1 & 1 & 1 \\
  0 & 0 & 1 & -1 & i & -i\\
  0 & 0 & 1 & -1 & -i & i\\
  0 & \sqrt{2} & 1 & 1 & 1 & 1\\
\end{array}%
\right)$ & $\frac{1}{2\sqrt{2}}\left(%
\begin{array}{cccccc}
  \sqrt{2} & \sqrt{2} & 1 & 1 & 1 & 1 \\
  0 & 0 & 1 & -1 & 0 & 0\\
  0 & 0 & 0 & 0 & -1 & 1\\
  \sqrt{2} & -\sqrt{2} & 0 & 0 & 0 & 0\\
\end{array}%
\right)$\\
  \hline
\end{tabular}
\end{table}

\section{\label{section-conclusions}Conclusions and Prospects}
\pst To conclude, we present the main results of the paper.

A bijective map: $\{N$ operators in $\mathcal{B}({H}_d)\} ~
\longleftrightarrow ~ \{ d^2\times N$ matrix $\mathscr{U}\}$ is
constructed and associated with a star-product formalism. For $N$
these operators to form a basis in $\mathcal{B}({H}_d)$,
conditions on matrix $\mathscr{U}$ are derived. Classification of
possible matrices $\mathscr{U}$ and related star-product schemes
$\{\hat{U}_k,\hat{D}_k\}_{k=1}^{N}$ is accomplished. This gives
rise to a new approach of introducing bases in
$\mathcal{B}({H}_d)$ with desired properties. One chooses a class
of matrices and impose additional limitations. Once matrix
$\mathscr{U}$ is built, a corresponding basis (set of operators)
in $\mathcal{B}({H}_d)$ with expected properties appears. A
development of the paper is complemented by illustrating examples.

Another substantial result is a series of Propositions and
Corollaries which demonstrate peculiarities of dequantizers and
quantizers, especially in a self-dual star-product scheme. Namely,
it is proved that there exists no minimal tomographic star-product
scheme with dequantizers in the form of POVM effects and Hermitian
semi-positive quantizers. On applying this argument to self-dual
schemes, we have proved that (i) there exists no minimal self-dual
star-product scheme with dequantizers in the form of POVM effects
and (ii) Hermitian dequantizers and quantizers of a self-dual
scheme must contain negative eigenvalues. The achieved results can
be useful for an analysis of the following problems which are of
great interest for further consideration: symmetric but
non-informationally complete structures of arbitrary rank, a
relation between symmetric bases in spaces of different dimension,
and specific bases in multipartite systems.

\section*{Acknowledgments}
\pst The authors thank the Russian Foundation for Basic Research
for partial support under Projects Nos. 09-02-00142 and
10-02-00312. S.N.F. is grateful to the Russian Science Support
Foundation for support under Project ``Best postgraduates of the
Russian Academy of Sciences 2010". S.N.F. thanks the Ministry of
Education and Science of the Russian Federation for support under
Project Nos. 2.1.1/5909, $\Pi$558, and 14.740.11.0497. V.I.M. was
supported by NIX Computer Company in the form of a gift (computer)
provided by Organizers of Seminars on quantum physics and
informatics in Landau Institute for Theoretical Physics, where
V.I.M. was happy to deliver two lectures.

\end{document}